 \documentclass[
 aps,
 amsmath,
 prd,
 twocolumn,
 groupedaddress,
 showpacs,
 nofootinbib
 ]{revtex4}

 \newcommand{\nn}{\nonumber}
 \newcommand{\td}[2]{\frac{{\rm d} {#1}}{{\rm d} {#2}}}
 \newcommand{\tdil}[2]{{\rm d} {#1} / {\rm d} {#2}}
 
 \newcommand{\pdil}[2]{\partial {#1} / \partial {#2}}

 \usepackage[dvips]{graphicx}
 \usepackage{amsmath}
 \usepackage{amsfonts}
 \usepackage{amssymb}

\begin{document}

 \title{The Mass of the Cosmos}

 \author{Charles Hellaby}
 \affiliation{Department of Mathematics and Applied Mathematics, \\
 University of Cape Town, Rondebosch 7701, South Africa}
 \email{cwh@maths.uct.ac.za}

 \date {19/11/2005}

 \begin{abstract}
   We point out that the mass of the cosmos on gigaparsec scales can be measured, 
owing to the unique geometric role of the maximum in the areal radius.
   Unlike all other points on the past null cone, this maximum has an associated 
mass, which can be calculated with very few assumptions about the cosmological 
model, providing a measurable characteristic of our cosmos.
   In combination with luminosities and source counts, it gives the bulk mass to 
light ratio.
   The maximum is particularly sensitive to the values of the bulk cosmological 
parameters.  
   In addition, it provides a key reference point in attempts to connect cosmic 
geometry with observations.
   We recommend the determination of the distance and redshift of this maximum be 
explicitly included in the scientific goals of the next generation of reshift surveys.
   The maximum in the redshift space density provides a secondary large scale 
characteristic of the cosmos.
 \end{abstract}

 \pacs{98.80.-k, 98.65.-r}

 \keywords{Cosmology, Cosmic Mass, Cosmic Density, Diameter Distance, Luminosity Distance, 
Redshify Space Density, Areal radius, Cosmic Parameters}

 \maketitle

 \section{Introduction}

   The emergence of automated large-scale redshift surveys \cite{SDSS,2dF} 
is opening up new possibilities 
for measuring the content and dynamics of the cosmos on very large scales.  We point 
out a significant characterisation of the cosmos on gigaparsec scales that will become 
measurable with the next generation of surveys.

   It is well known in observational cosmology that our past null cone (pnc) has a maximum in 
its areal radius, $\hat{R}_m$, where the angular size of sources of a given size is minimum.  
Beyond this point, more distant images, though dimmer, subtend larger angular sizes 
\cite{Hoyl61,McC34}.  It is also known in relativistic cosmology that this maximum occurs 
where the observer's pnc crosses the apparent horizon.  What hasn't been realised 
is that a measurement of this maximum is equivalent to a measurement of the mass within a 
sphere of areal radius $\hat{R}_m$, and that this relationship is quite general, not requiring 
the assumption of homogeneity for example.
   Less well known is that the number density of sources in redshift space also has a 
maximum.  This latter maximum, however, does not have such a deep significance as the former.

   It is best to use the 
 Lema\^{\i}tre-Tolman (LT) model, as both $R$ and $M$ are primary functions, so the 
relationships we are interested in are particularly clear.  Also, the equation of state
 --- pressure free matter plus $\Lambda$
 --- is very suitable for the post-recombination universe, and the spherical symmetry 
of the model about the origin is entirely natural in the context of observations, since 
isotropy is fairly well established, and of course our own past light cone is centered 
on ourselves.

   The Lema\^{\i}tre-Tolman (LT) metric \cite{Lem33,Tol34} is
 \begin{align}
   ds^2 = - dt^2 + \frac{(R')^2}{1 + 2 E} \, dr^2
   + R^2 (d\theta^2 + \sin^2\theta \, d\phi^2)
   \label{LTmetric}
 \end{align}
 where $R = R(t, r)$ is the areal radius, $R' = \pdil{R}{r}$, and $E = E(r)$ is an arbitrary 
function of coordinate radius $r$.  We use geometric units in all equations.  This metric describes 
 pressure-free matter in comoving coordinates.  From the Einstein field equations,
 \begin{align}
   \dot{R}^2 & = \frac{2 M}{R} + 2 E + \frac{\Lambda R^2}{3} ~,
   \label{EvEqLT} \\
   \kappa \rho & = \frac{2 M'}{R^2 R'} ~,
   \label{DensityLT}
 \end{align}
 where $\kappa = 8 \pi G/c^4 = 8 \pi$, $\rho(t,r)$ is the density, 
$\dot{R} = \pdil{R}{t}$ and $M = M(r)$ is also arbitrary.  The solutions to 
(\ref{EvEqLT}) are best found numerically for the general case.  The density is divergent 
at the bang and crunch, $R = 0$, and also at shell crossings, where $R' = 0$ but $M' \neq 0$.  
The latter singularity is avoidable \cite{HelLak85}, but the former isn't.

   As can be seen from (\ref{LTmetric}), the function $E$ determines the local geometry.  
In addition, (\ref{EvEqLT}) shows that it is a kind of energy parameter.  Function $M$ 
is the gravitational mass within a comoving sphere of radius $r$, so it includes any putative 
dark matter component(s), but does not include the ``density" associated with $\Lambda$.  
This $M$ is not the same as the integrated proper density $\cal M$,
 \begin{align}
   {\cal M} = \int_0^r \rho \, d^3V
      = \int_0^r \big( M' / \sqrt{1 + 2 E}\; \big) \, dr ~.
   \label{IntegratedDensity}
 \end{align}
 This latter is the mass one would obtain by summing the masses of individual galaxies, gas 
clouds, dark matter concentrations, etc.  See \cite{Kra97} for further discussion of the LT 
model and supporting references.

 \section{Observables}

   The light rays making up the pnc (past null cone) are the incoming radial 
null geodesics arriving at the central observer at a given moment, 
in particular:
 \begin{align}
   \td{t}{r} = \frac{- R'}{\sqrt{1 + 2E}\;}
   ~,~~~~~~ t = t_0 ~~\mbox{at}~~ r = 0
      \label{tn_DE_LT}
 \end{align}
 with solution $t = \hat{t}(r)$.  Any quantity $Q(t,r)$ evaluated on our current 
pnc will be indicated with a hat: $\hat{Q} = \hat{Q}(r) = Q(\hat{t}(r),r)$, 
and for expressions a square bracket with subscript ``$n$" will be used.  
The redshift of an observed object, located at $r_e$, is given by
 \begin{align}
   \ln(1 + z) = \int_0^{r_e} \frac{\widehat{\dot{R}'}}{\sqrt{1 + 2 E}\;} \, dr
      \label{zLT}
 \end{align}
 where $\dot{R}'$ may be found from (\ref{EvEqLT}).  
 Along a ray, the areal radius is $\hat{R} = R(\hat{t}(r), r)$, 
 and its rate of change along the ray, using (\ref{EvEqLT}) and (\ref{tn_DE_LT}), is
 \begin{align}
   \hat{R}' & = \left[ \dot{R} \hat{t}' + R' \right]_n \nonumber \\
      & = \left[ \left( -
      \frac{\sqrt{\frac{2 M}{R} + 2E + \frac{\Lambda R^2}{3}}\;}{\sqrt{1 + 2E}\;}
      + 1 \right) R' \right]_n.
   \label{AH_eq_LT}
 \end{align}
 Since the coordinate $r$ is not observable or physically meaningful we calculate 
 \begin{align}
   & \td{\hat{R}}{z} = \frac{\hat{R}'}{z'}
      = \Bigg[ \frac{R'}{\dot{R}' (1 + z)} \Bigg( \sqrt{1 + 2E}\; \nn \\
   &~~~~ - \sqrt{\frac{2 M}{R} + 2E + \frac{\Lambda R^2}{3}}\;  \Bigg) \Bigg]_n ~.
   \label{dRndz_LT}
 \end{align}
 Near the origin we have $M \to 0$, $E \to 0$ and $z \to 0$, so
 \begin{align}
   \tdil{\hat{R}}{z} \to [ R' / \dot{R}' ]_{n\,0} ~.
   \label{dRndz_LT_0}
 \end{align}
 See \cite{MuHeEl97,MuBaHeEl98} for further details.


   If a source such as a galaxy has measured angular diameter $\delta$ and known actual 
diameter $D$, then the diameter distance is identically the areal radius on the pnc
and is defined by
 \begin{align}
   \hat{R} = d_D = D / \delta ~.
   \label{DiamDistDef}
 \end{align}
   If the apparent luminosity of an observed source is $\ell$, and its absolute 
luminosity is known to be $L$, then the luminosity distance is defined by
 \begin{align}
   d_L = \sqrt{L / \ell}\; d_a
   = 10^{(m - \tilde{m})/5} \, d_a ~,
   \label{LumDistDef}
 \end{align}
 where $d_a = 10$~pc, $m$ is the apparent magnitude and $\tilde{m}$ the absolute 
magnitude.  For a general curved spacetime and arbitrary 
motion, the reciprocity theorem tells us
 \begin{align}
   d_L = d_D(1 + z)^2 ~.
   \label{ReciprocityThm}
 \end{align}
 If $d_D = \hat{R}$ is calculated from $m$ \& $z$ measurements using (\ref{ReciprocityThm}), 
the fractional error is comparable with that of $d_L$, assuming the redshifts are fairly 
accurate:
 \begin{align}
   \delta \hat{R} / \hat{R} & = 0.2 \, \ln 10 \, \delta (m - \tilde{m})
      - 2 \delta z / (1 + z) ~.
 \end{align}
For each of these distances, it is essential to know an intrinsic source property, 
$D$ or $L$ (or $\tilde{m}$).  While their values for nearby sources can be determined 
using other distance measures, for distant sources there is the extra difficulty 
of determining how much they have evolved with time, or even what type of nearby 
object the source corresponds to.

   If $n$ is the number density of sources in redshift space(/steradian/unit redshift 
interval), and $\mu$ is the mass per source, then the relation between $n$ and $\rho$ 
is
 \begin{align}
   \mu n & = \Bigg[ \frac{\rho R^2 R'}{\sqrt{1 + 2 E}\;} \Bigg]_n \, \frac{dr}{dz}
   = \Bigg[ \frac{2 M'}{\kappa \dot{R}' (1 + z)} \Bigg]_n
   = \Bigg[ \frac{2}{\kappa} \frac{d{\cal M}}{dz} \Bigg]_n
 \label{mun}
 \end{align}
 See \cite{Hel01} for a discussion of multiple source types and multicolour observations.

 \section{Characteristic Cosmic Mass and Density}
 \label{CCMD}

   In an expanding, non-inflating universe, the diameter distance $\hat{R}(r)$ 
necessarily has a maximum.  
 The locus of such points, where different rays are momentarily at constant $R$, 
is the apparent horizon (AH).  We find it by putting $\hat{R}' = 0$ in (\ref{AH_eq_LT}), 
giving
 \begin{align}
   \Lambda \hat{R}_m^3 - 3 \hat{R}_m + 6 M_m & = 0 ,
   \label{AH_cubic_LT}
 \end{align}
 so thus if $R_m$ is on the AH, then $2 M_m = R_m - \Lambda R_m^3 / 3$, and if 
$\Lambda = 0$ this is the familiar $R = 2 M$.  
   In general the locus of the pnc, and the variation of measurables down 
it, are strongly affected by the details of the cosmological model.  This point, 
however, where the pnc crosses the AH, has a unique meaning:
 \begin{itemize}
 \item[(a)] there is a simple direct relationship between $R$ and $M$ that is 
independent of any inhomogeneities in $E$, $t_B$ and $M$,
 \item[(b)] since $R$ is measurable, the mass $M$ within that radius is 
immediately determined,
 \item[(c)] this point flags a major causal feature of model,
 \item[(d)] the maximum in $R$ is a distinctive feature of the $R$-$z$ plot.
 \end{itemize}
 No other point on the pnc has such a simple, generic $M$-$R$ relationship, and this 
holds true whether not the model is homogeneous.  Therefore $R_m$ and the redshift 
$z_m$ where it occurs are distinctive characteristics of the 
cosmological model.

   For $M_m \geq 0$ \& $R_m \geq 0$, equation (\ref{AH_cubic_LT}) has solutions if 
 \begin{align}
   M_m \leq 1 / (3 \sqrt{\Lambda}\;) = (M_m)_{max} ~,
   \label{Mm_max_LT}
 \end{align}
 and this maximum possible value of $M_m$ occurs at 
 \begin{align}
   (R_m)_{max} = 1 / \sqrt{\Lambda}\; ~.
   \label{RofMm_max_LT}
 \end{align}
 Thus if an incoming ray reaches $R > (R_m)_{max}$, then it has no maximum in $R$.  
Worldlines with $M > (M_m)_{max}$ never meet the AH, and those with $M < (M_m)_{max}$ 
cross it twice.  

   Condition (\ref{AH_cubic_LT}) will always hold somewhere for any cosmology with a 
 non-zero matter density, because $M = 0$ at the observer, and increases outwards.  
In all LT models with a bang, the AH goes as $\hat{R} \approx 2 M$ near the bang.  In 
 ever-expanding models, the AH asypmtotically approaches the de Sitter horizon, 
$R = \sqrt{3/\Lambda}\;$, as $t \to \infty$.  The ``incoming" ray that is tangential 
to $R = \sqrt{3/\Lambda}\;$ at $t = \infty$ divides rays that have a maximum and reach 
the origin, from those that have no maximum and have
 ever-increasing $\hat{R}$.  In closed 
 re-collapsing models, putting $\dot{R} = 0$ and $E = -1/2$ in (\ref{EvEqLT}) reproduces 
(\ref{AH_cubic_LT}), thus showing that the moment of maximum expansion at the maximum in 
the spatial sections lies on the AH.  (In fact it is where the past and future AHs 
cross.)  Thus all pncs in (physically well-behaved) closed 
 re-collapsing LT models have maxima in their areal radii.  For further discussion of 
the apparent horizon, see 
\cite{Hel87,KraHel04b}.

   It is evident from (\ref{mun}) that the redshift space density also has a maximum, 
because, although $M' \geq 0$ and increases from zero at the origin%
 \footnote{Here we use a well behaved $r$ coordinate, such as one with $M \propto r^3$ 
or $R(t = {\text const}, r) \propto r$.}%
 , both $\dot{R}'$ and $(1 + z)$ are finite at the observer, and diverge towards the 
bang.  For the maximum in $\mu n$, we put 
 \begin{align}
   & 0 = \frac{d(\mu n)}{dz}
      = \Bigg[ \frac{2}{\kappa} \frac{d^2{\cal M}}{dz^2} \Bigg]_n
      = \Bigg[ \frac{2}{\kappa (\dot{R}')^3 (1 + z)^2} \times \nn \\
   & \Big\{\sqrt{1 + 2 E}\; (M'' \dot{R}' - M' \dot{R}'')
      + M'(\ddot{R}' R' - (\dot{R}')^2) \Big\} \Bigg]_n ,
 \label{dmundz}
 \end{align}
 and solve for $\widetilde{\mu n}_m$ and $\tilde{z}_m$.  The locus of this maximum 
doesn't have a deep geometric or physical meaning, but depends on the redshift 
behaviour down the pnc, which depends strongly on the details of the model.  
Nevertheless, one can say that the maximum in $\mu n$ is a large scale characteristic 
density of the model.

 \section{Robertson-Walker Case}

   The key results are those given above in section \ref{CCMD}, which apply to 
a fairly general class of realistic
 post-recombination cosmologies.  However, it is useful to look at the 
characteristic cosmic mass and densitiy in the standard homogeneous model.  The 
above results are specialised to the homogeneous case in appendix \ref{RWap}.

   The Robertson-Walker (RW) version of the AH equation (\ref{AH_cubic_RW}) seems quite 
complex, and hides the key relationship that is evident in (\ref{AH_cubic_LT}).

\begin{figure}[h]
    \includegraphics[scale=0.4]{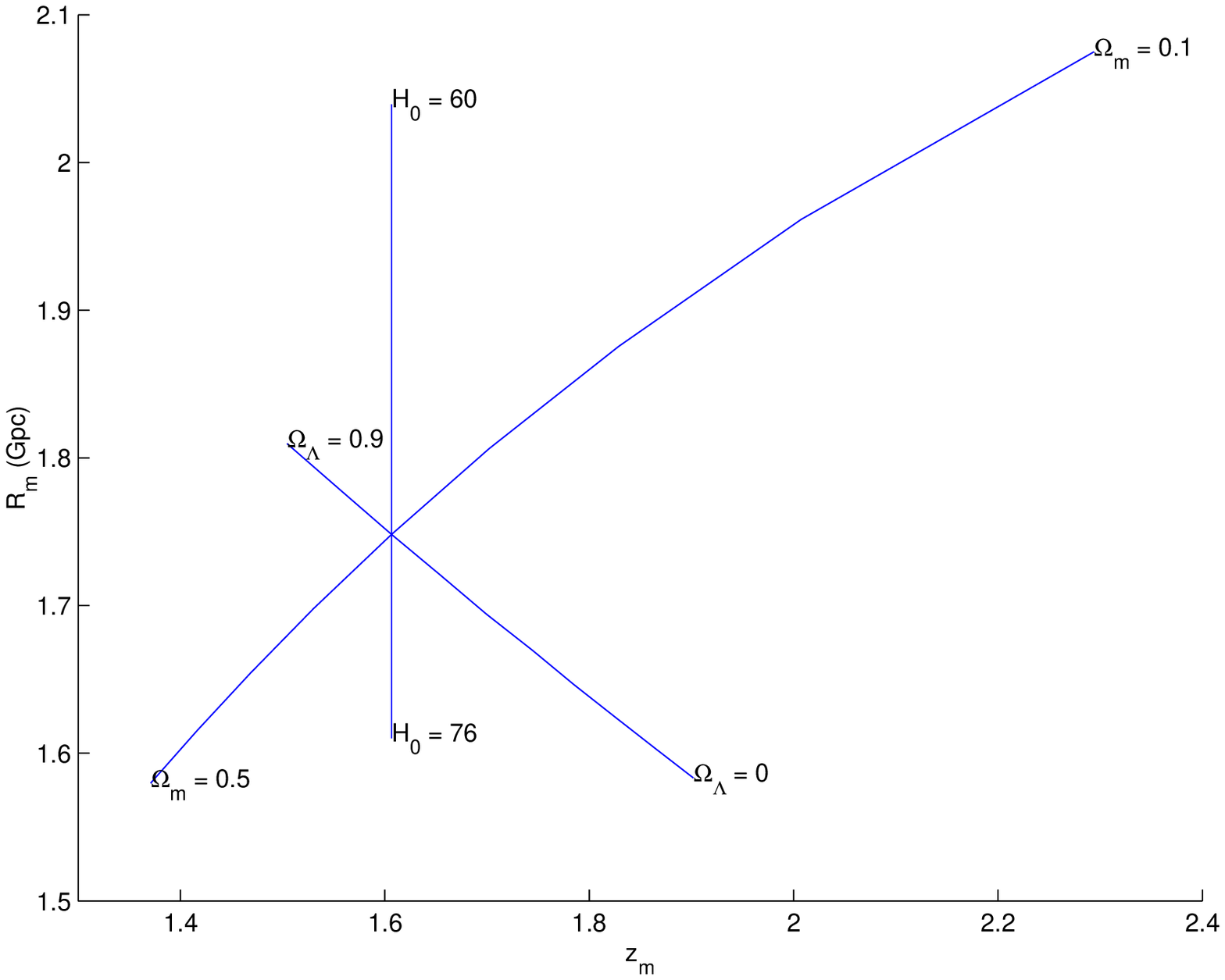}
    \includegraphics[scale=0.4]{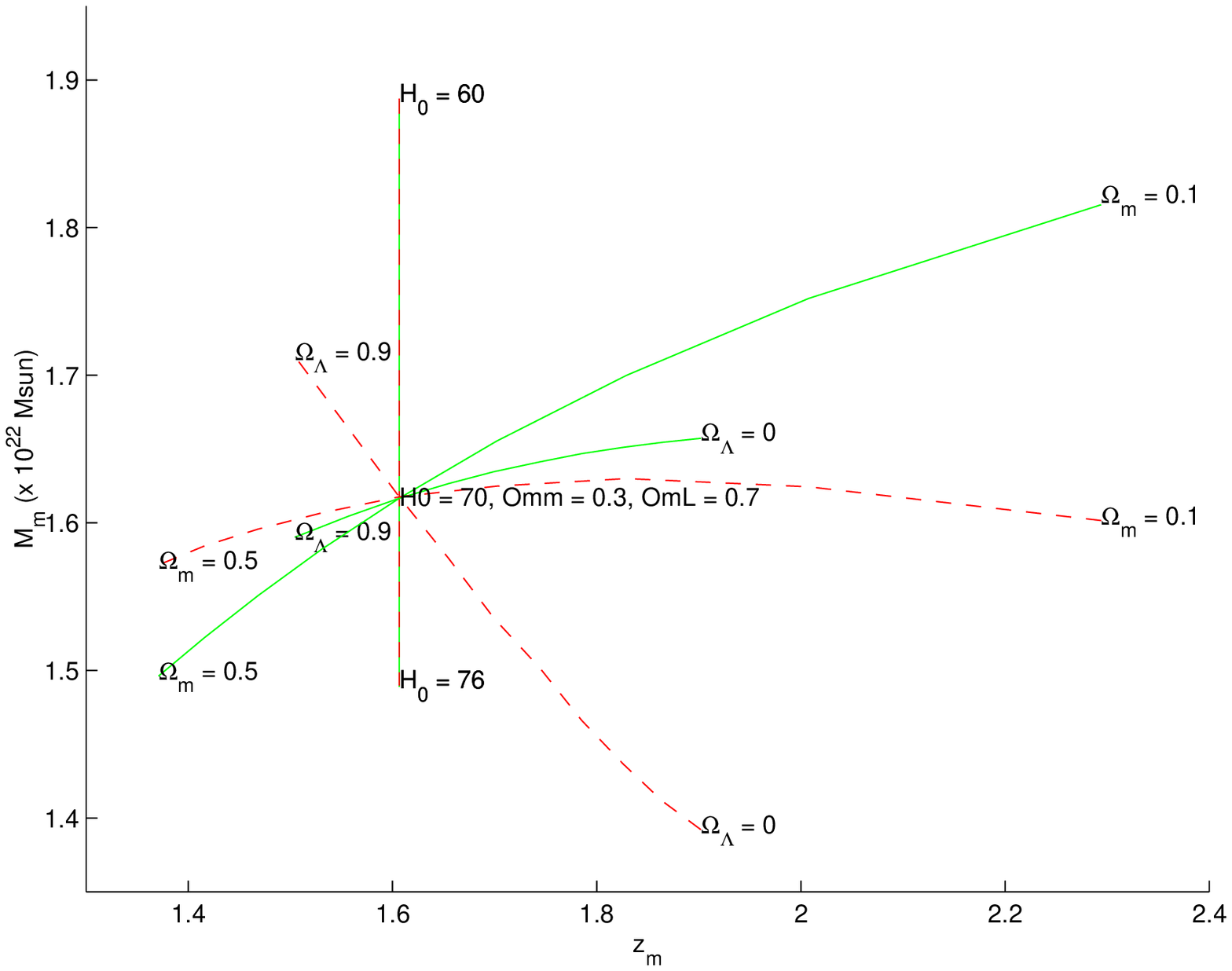}
    \includegraphics[scale=0.4]{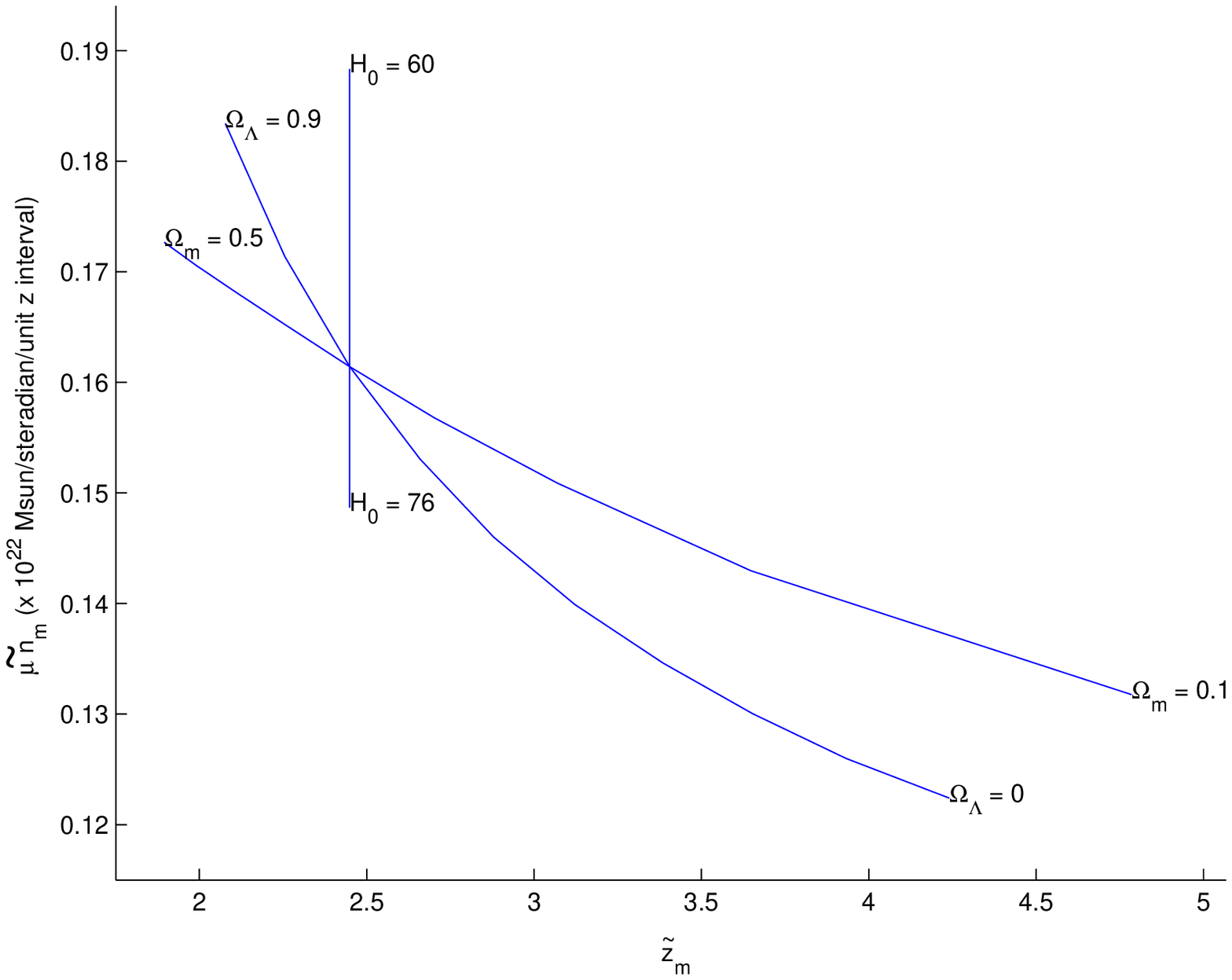}
 \caption{
 \label{RmaxMmaxVariation}
 (a) The dependence of $\hat{R}_m$ and $z_m$ on the RW parameters; 
 (b) the dependence of $M_m$ (solid lines) and ${\cal M}_m$ (dashed lines) 
on the RW parameters.
 (c) the dependence of $\widetilde{\mu n}_m$ and $\tilde{z}_m$ on the RW parameters.
 }
 \end{figure}

The dependence of $\hat{R}_m$, $M_m$, ${\cal M}_m$ and $\widetilde{\mu n}$ on $H_0$, 
$\Omega_\Lambda$ and $\Omega_m$ is shown in fig \ref{RmaxMmaxVariation}, and for 
reference the $\hat{R}(z)$ and $\mu n(z)$ curves for a range of values of these 
parameters are given in the appendix%
 \footnote{
 For a related earlier treatment with $\Lambda = 0$, see \cite{EllTiv85}.
 }%
 .

 \section{Discussion}

   The properties outlined in section \ref{CCMD} below eq (\ref{AH_cubic_LT}) make the 
maximum in $\hat{R}$ a very significant feature of a cosmological model, both 
theoretically and observationally.

   Both $\hat{R}_m$ and $z_m$ may be measured, and calculating $M_m$ from 
(\ref{AH_cubic_LT}) requires only the value of $\Lambda$, and is independent of 
the details of the cosmology, or even whether large-scale homogeneity exists.  
Note also that $M_m$ is not very sensitive to uncertainties in $\Omega_\Lambda$; for 
a given measured $R_m$, an increase from $0$ to $0.7$ decreases the calculated $M_m$ 
by $12$\%%
 \footnote{
 This is the uncertainty in $M_m$ once $R_m$ has been measured, rather than the 
variation in the $M_m$ predicted by a variety of models with different 
$\Omega_\Lambda$ values, in which $R_m$ also varies.  
 }%
 .

   Current galaxy redshift surveys only extend to $z = 0.3$, while quasar redshift 
surveys extend to $z = 6$.  However the data 
is not complete enough, and the quasar population too diverse to be useful.  
The next generation of galaxy surveys will extend past the redshift of the maximum, 
and recent supernova observations \cite{StroEtAl04} are already approaching it.  
   The $\hat{R}$-$z$ plot may be obtained from direct measurements of redshifts and 
angular diameters, or derived from $m$-$z$ measurements.  While the former would be 
ideal, the latter is acceptable, since the reciprocity theorem is very general.  
Of course, knowledge (or assumptions) about true diameters and absolute luminosities 
of sources, and their $z$ evolution, is essential, and at large $z$ this is a 
significant uncertainty.

   While locating $\hat{R}_m$ reliably requires a good sample of data points in a 
range near $z_m$, determination of $n(z)$ is not so easy, since one must be sure of 
detecting or reliably estimating all masses.  Nevertheless, a sufficiently deep, 
complete survey should provide an indication of $\widetilde{\mu n}_m$ and $\tilde{z}_m$.

 An issue for future consideration is the effect of inhomogeneities on the 
uncertainty in $\hat{R}_m$ and $z_m$ determinations.  For example \cite{MuBaHeEl98} 
showed inhomogeneity can cause loops in the $\hat{R}$-$z$ curve near the maximum.

   The 
 mass-to-light ratio is difficult to determine outside gravitationally 
bound systems, and direct estimates of (gravitational) mass from galaxy and cluster 
dynamics, and from gravitational lensing only extend up to cluster or supercluster 
size, whereas the determination of $M_m$ reaches several orders of magnitude larger.  

   The possibility of determining the metric of the cosmos from observations of 
redshifts, luminosities (or angular diameters) and the number density of sources, 
combined with the evolution of absolute luminosities, true diameters, and mass per 
source, was considered in \cite{ObsCosmol, MuHeEl97, Hel01}.  
 A project to begin implementing this is now underway \cite{LuHel05}.  
 The cosmic mass $M_m$ provides an important 
 cross-check on the summed mass at that radius.  In fact the theorem of \cite{MuHeEl97} 
 --- that, given any reasonable set of observations and any reasonable source evolutions 
functions, an LT model can be found that fits them
 --- needs to be qualified, as the combination of observations and evolution functions 
must mesh correctly close to $\hat{R}_m$.

   Within the 
 dust-$\Lambda$-RW model, it is noteworthy that the maximum is where the 
$\hat{R}(z)$ curve is most sensitive to variations in $\Omega_\Lambda$ and $H_0$, 
and nearly so for $\Omega_m$, and that variations in these 3 parameters move the 
$(z_m, \, \hat{R}_m)$ and $(\widetilde{\mu n}_m, \, \tilde{z}_m)$, loci in very 
different directions.  Thus a determination of $\hat{R}_m$ \& $z_m$ would provide 
rather generic limits on $H_0$, $\Omega_m$ and $\Omega_\Lambda$, and combined with 
the initial slope of the $\hat{R}$-$z$ graph, or with measurements of 
$\widetilde{\mu n}_m$ and $\tilde{z}_m$, would fix all three values.  Determining 
$\hat{R}_m$ and $z_m$ to within 10\% would by itself provide confirmation (or 
otherwise) of current parameter estimations, while 5\% accuracy would put new 
constraints on the possible values.

   The problem of averaging in GR means that identifying the RW model that best fits the 
observations is not a well-defined execise.  Therefore measurments of bulk effects are 
particularly important, and since $R_m$, unlike any other point on the pnc, gives the 
total mass on that scale, the $R_m$ and $z_m$ values provide a natural definition of the 
best fit RW model.

   Although particular models always have a $\hat{R}$-$M$ relation, such as 
(\ref{RW_M_E_tB}) \& (\ref{RW_R_SEv}) would give for RW models, this relation is very 
model dependent, whereas $M_m$ is not.  It is more general even than the 
LT model, as any cosmology will have a locus where the past null cone crosses the 
apparent horizon, and an associated mass is naturally defined there.

 \section {Conclusion}

   In summary, the maximum in the diameter distance is the only point on the past null 
cone that corresponds to a model independent mass, thus allowing direct measurement of 
a characteristic cosmic mass on gigaparsec scales.
  Therefore it also provides a very 
 large-scale check on the mass to light ratio, as well as a reference point for 
determining geometry from observations.  Since it is a point on the apparent 
horizon, a measurement of this maximum may actually be the first detection of a 
relativistic horizon.

   For RW models, the region near the maximum is where $R(z)$ is most sensitive to the 
values of the RW parameters.

   We advocate that, with the next generation of surveys, direct measurements of 
angular sizes and hence diameter distances be compiled and calculated independently 
of luminosity distances, and that, apart from fitting a 
Friedmann-Lema\^{\i}tre-Robertson-Walker (FLRW) model to the available 
data, a separate determination of $\hat{R}_m$ be done with a limited data set near 
$z_m$, thus giving a model independent value for $M_m$.

 \begin{acknowledgements}
   CH thanks the South African National Research Foundation for a grant.
 \end{acknowledgements}
 \vspace*{-2mm}

 \begin{figure}[h]
    \includegraphics[scale=0.4]{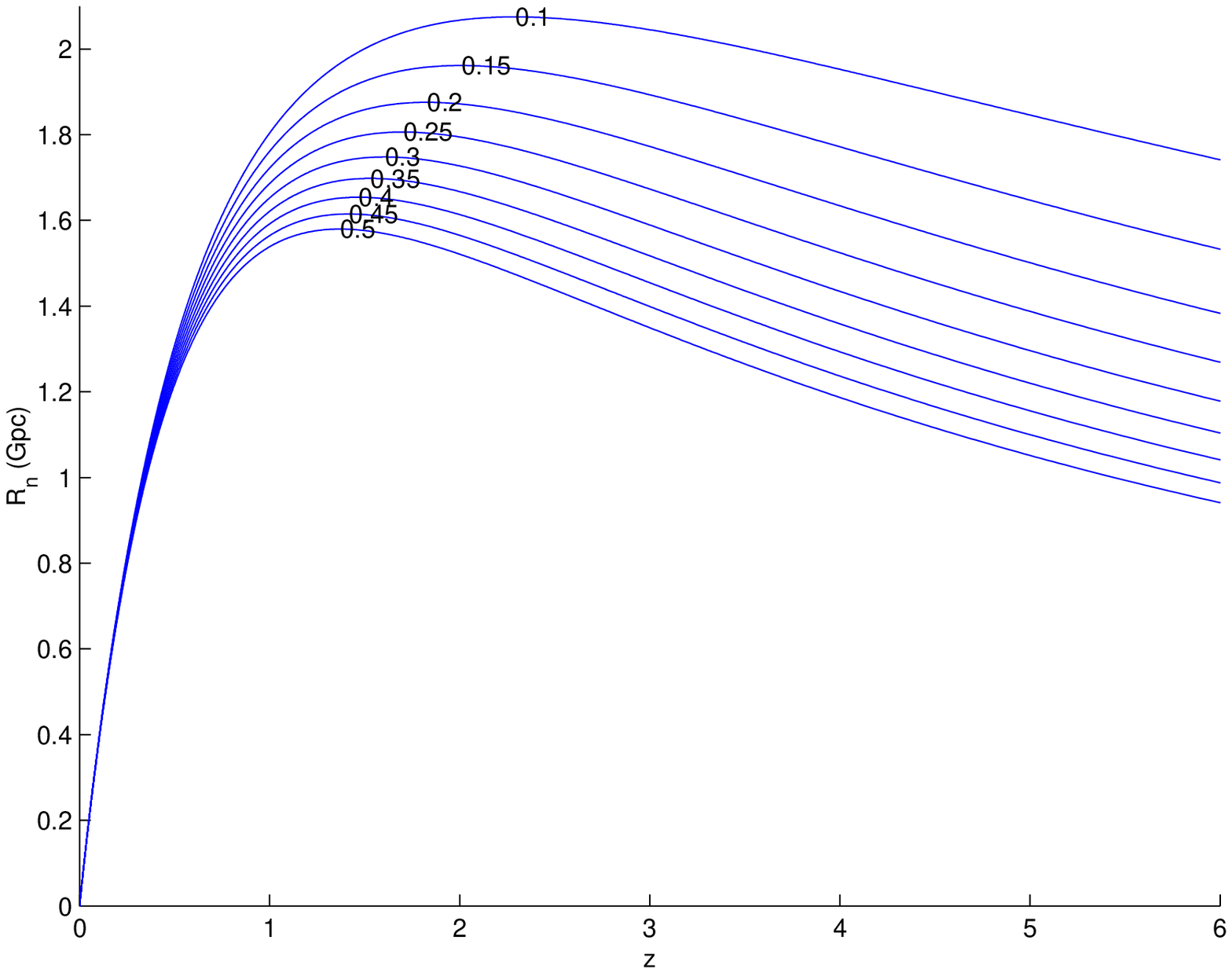}
    \includegraphics[scale=0.4]{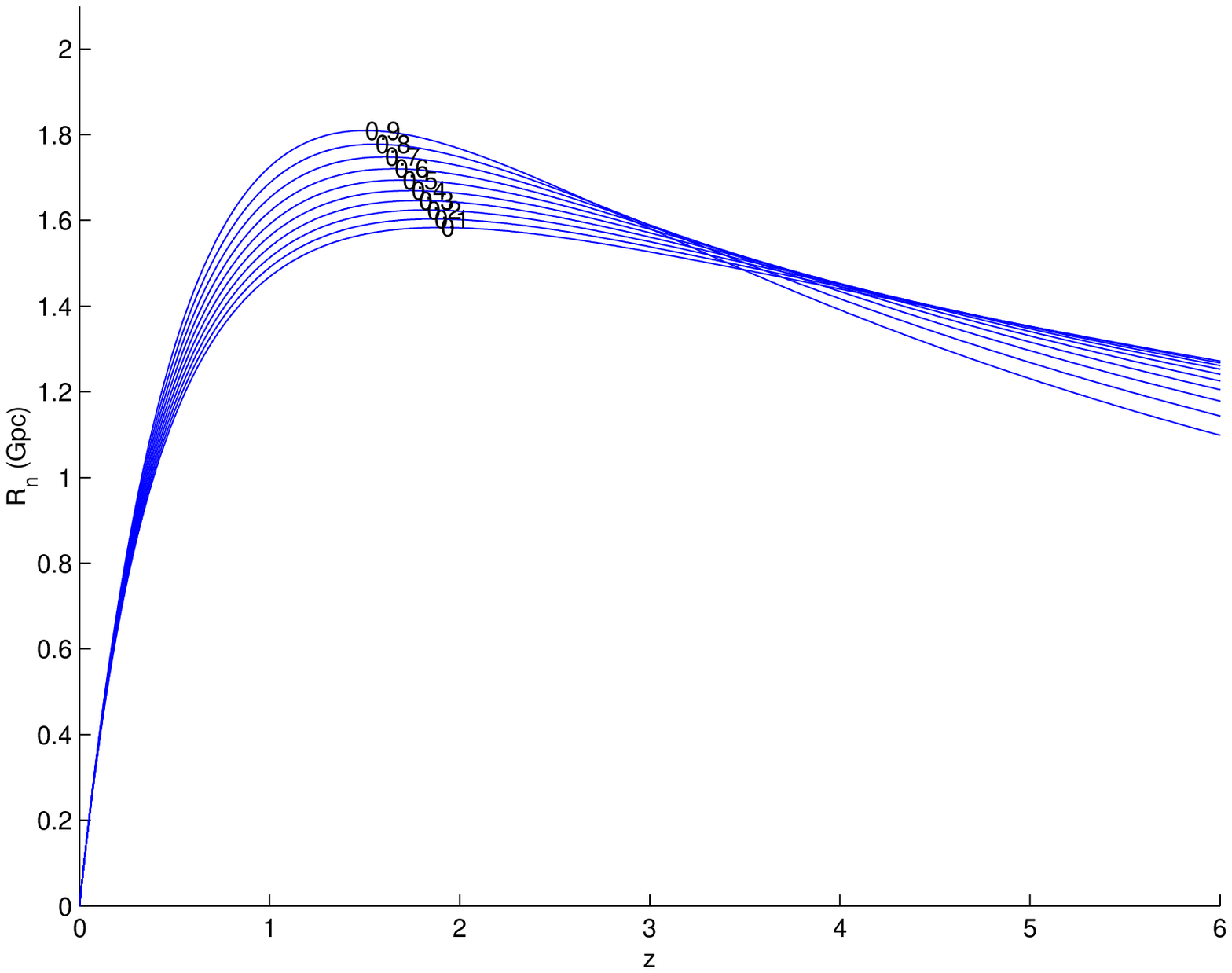}
    \includegraphics[scale=0.4]{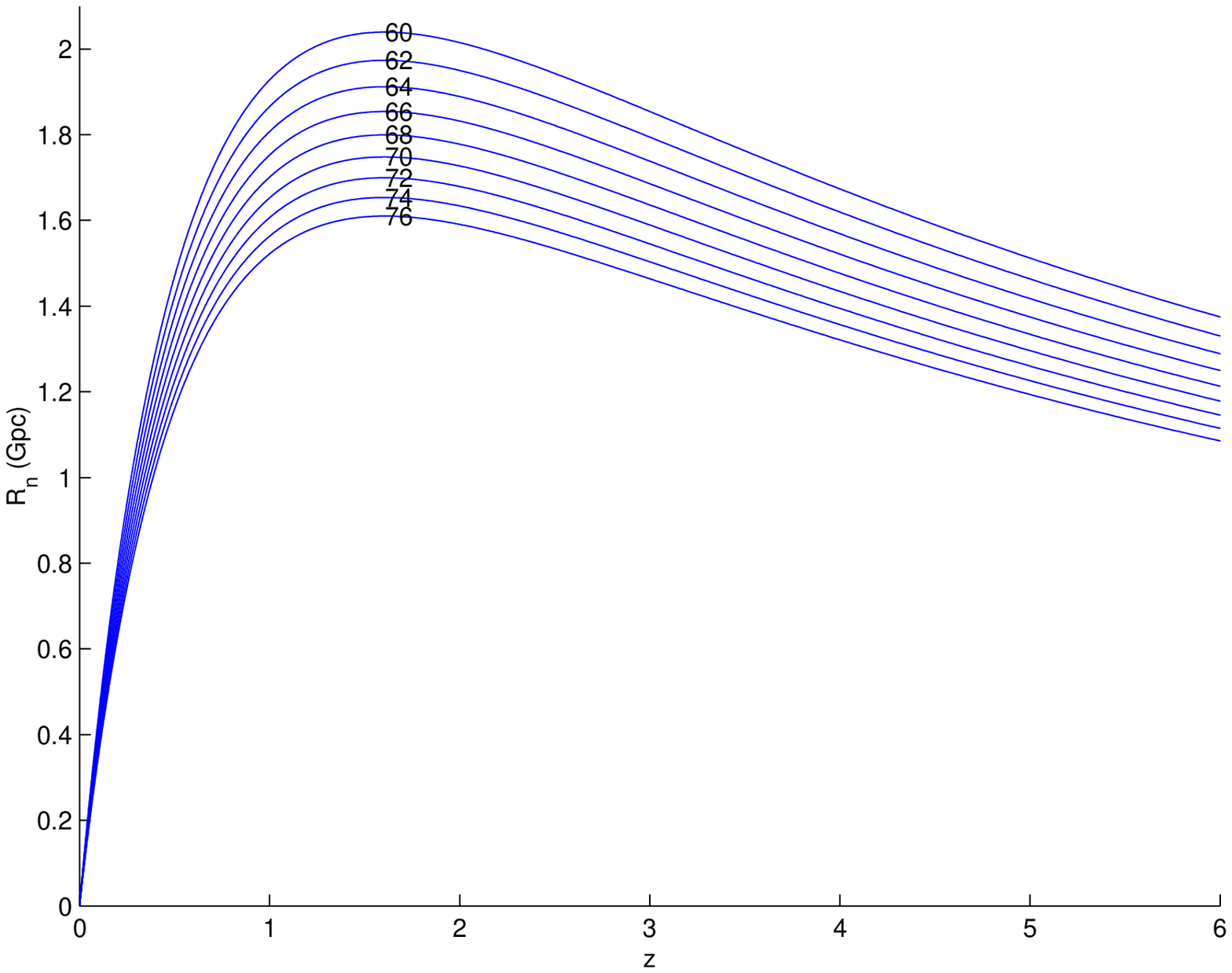}
 \caption{
 \label{R-zCurves}
 The $\hat{R}(z)$ curves for a range of RW parameters: 
 (a) different $\Omega_m$ values ($H_0 = 70$, $\Omega_\Lambda = 0.7$),
 (b) $\Omega_\Lambda$ values ($H_0 = 70$, $\Omega_m = 0.3$),
 (c) $H_0$ values ($\Omega_m = 0.3$, $\Omega_\Lambda = 0.7$). 
 }
 \end{figure}

 \begin{figure}[h]
    \includegraphics[scale=0.4]{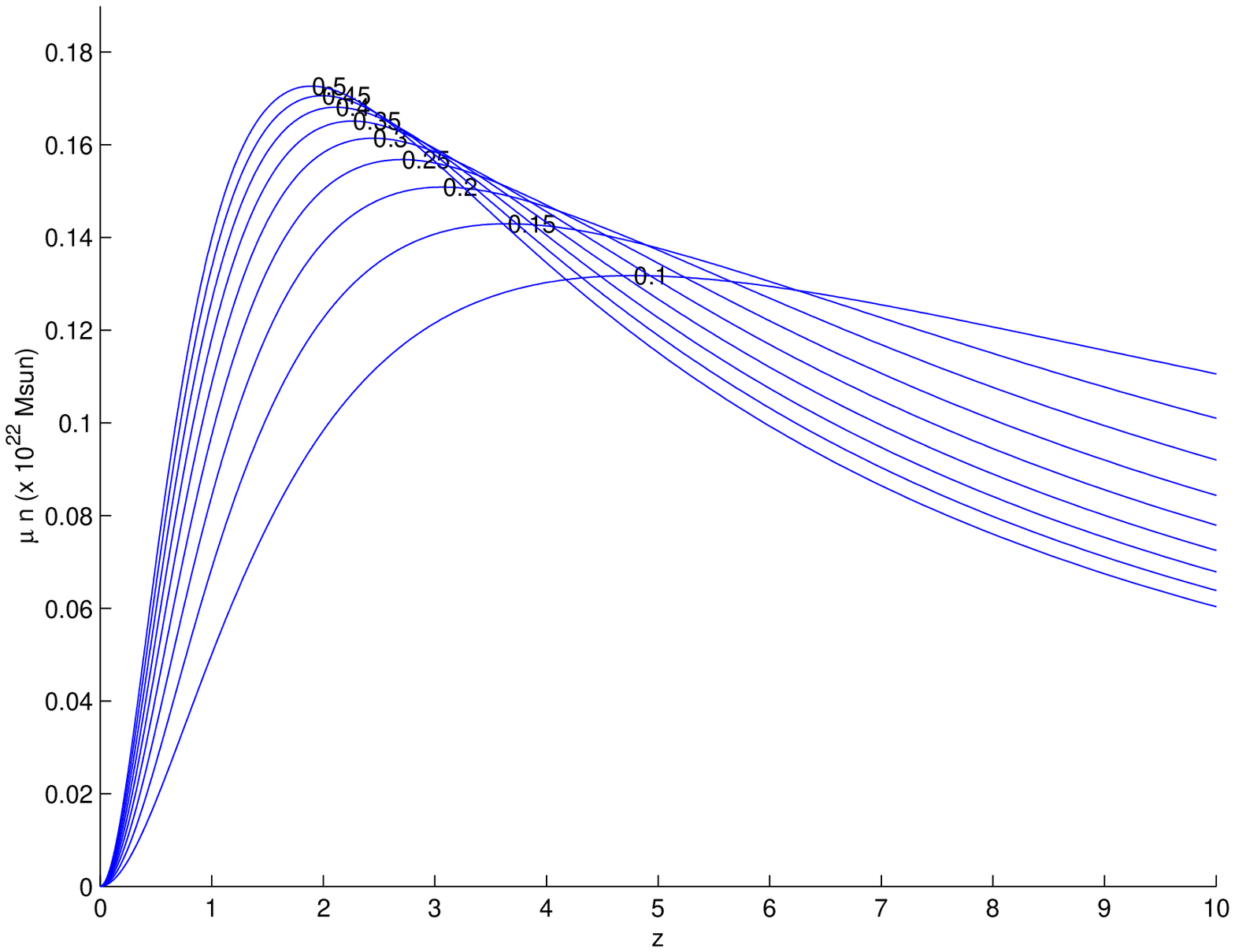}
    \includegraphics[scale=0.4]{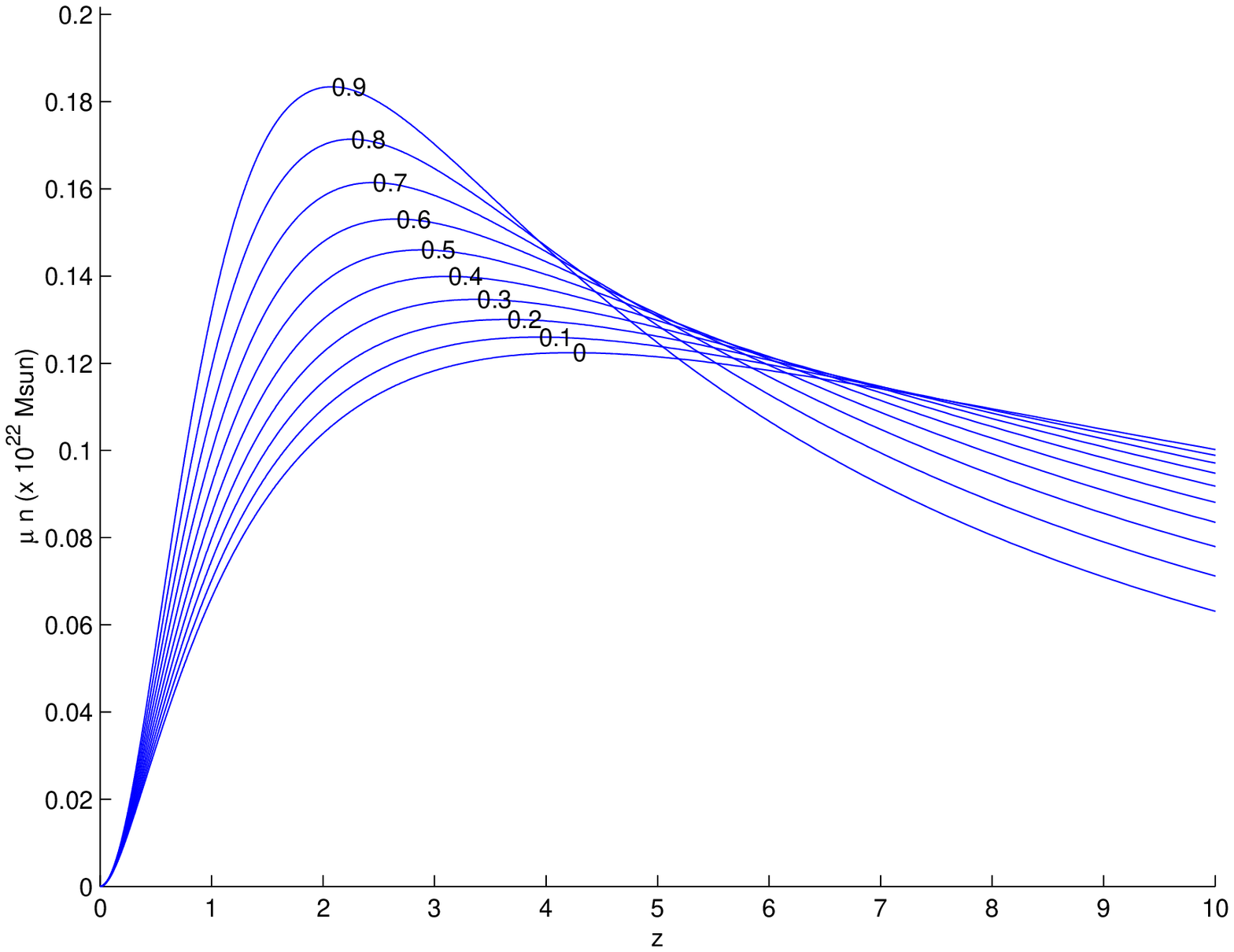}
    \includegraphics[scale=0.4]{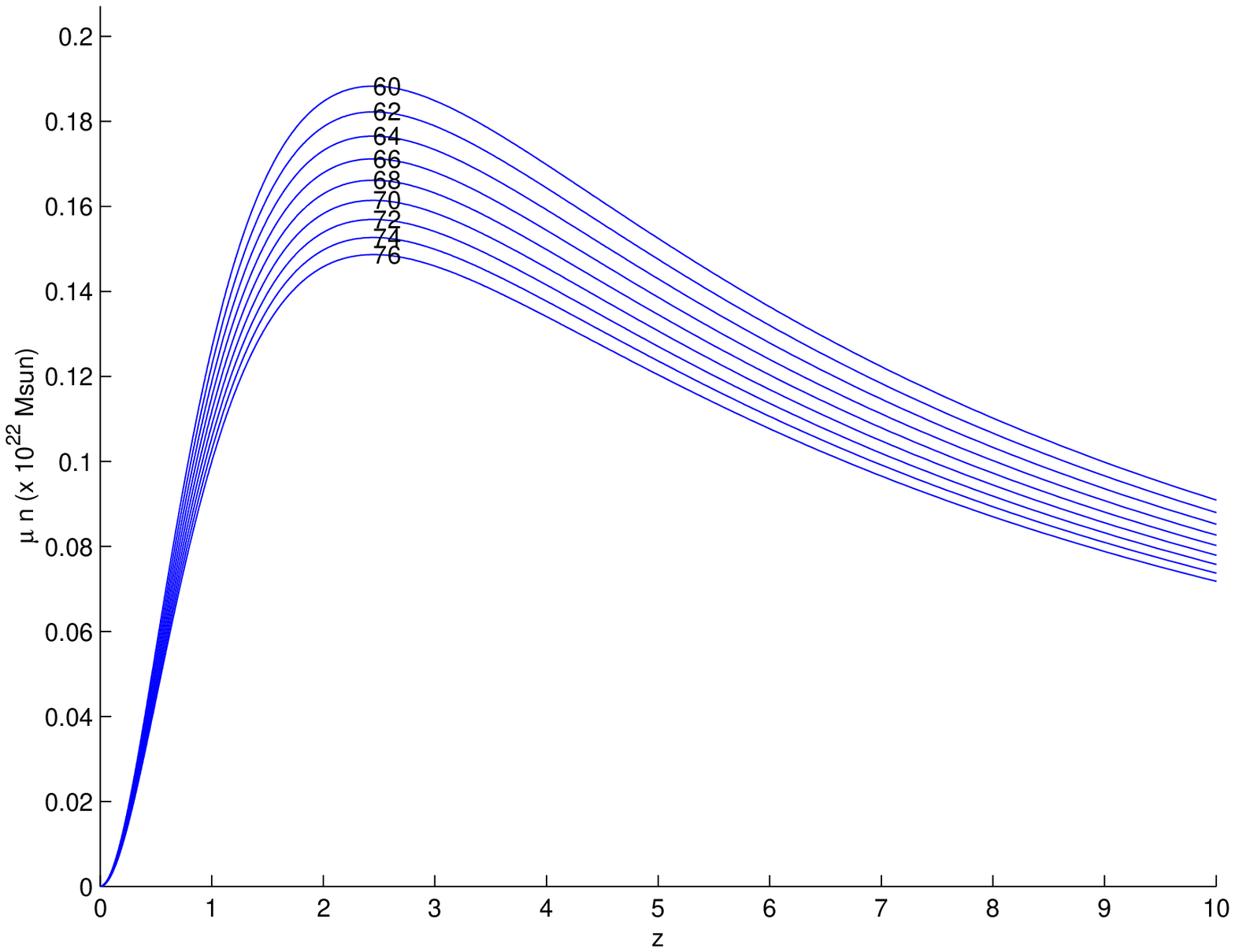}
 \caption{
 \label{mun-zCurves}
 The $\mu n(z)$ curves for a range of RW parameters (note that the $z$ scale 
is different from that of fig. (\ref{R-zCurves}): 
 (a) various $\Omega_m$ values ($H_0 = 70$, $\Omega_\Lambda = 0.7$),
 (b) $\Omega_\Lambda$ values ($H_0 = 70$, $\Omega_m = 0.3$),
 (c) $H_0$ values ($\Omega_m = 0.3$, $\Omega_\Lambda = 0.7$). 
 }
 \end{figure}

 \appendix

 \section{Specialising the LT equations to RW}
 \label{RWap}

   The
 dust-$\Lambda$-RW model, in its most common coordinate system, is obtained if 
we put
 \begin{align}
   M = M_0 r^3 ~,~~~~
   2 E = - k r^2 ~,~~~~
   t_B = 0 ~,
 \label{RW_M_E_tB}
 \end{align}
 from which we have
 \begin{align}
   R = r S(t) ~,~~~~~
      \dot{S}^2 = 2 M_0 / S - k + \Lambda S^2 / 3 ~,
      \label{RW_R_SEv} \\
   H_0 = \dot{S}_0 / S_0 ~,~~~
      \kappa \rho = 6 M_0 / S^3 ~,~~~
      \kappa \rho_0 = 6 M_0 / S_0^3 ~,
 \end{align}
 where $S$ is the scale factor.  
 Given $H_0$, $\Omega_\Lambda$ and $\Omega_m$, then
 \begin{align}
   t_0 & = 2 / 3 H_0 ~,~~~~
      \kappa \rho_0 = 3 \Omega_m H_0^2 ~,~~~~
      \Lambda = 3 H_0^2 \Omega_\Lambda ~, \\
   \Omega_k & = 1 - \Omega_\Lambda - \Omega_m ~,~~~~
      k = - {\rm sign}(\Omega_k) ~, \\
   S_0 & = \left\{ \begin{array}{ll}
           \text{arbitrary} & ~\mbox{if}~~ k = 0 ~, \\
           \frac{1}{H_0} \sqrt{\frac{-k}{\Omega_k}}\; & ~\mbox{if}~~ k \neq 0 ~,
           \end{array} \right. \\
   M_0 & = \Omega_m H_0^2 S_0^3 / 2 ~,
 \end{align}
 where $\Omega_m$ is the baryonic plus dark matter fraction.  
 The integrated density equation is
 \begin{align}
   {\cal M} & = \frac{3 M_0 k}{2} \Bigg( \frac{\sin^{-1}(\sqrt{k}\; r)}{\sqrt{k}\;}
      - r \sqrt{1 - k r^2}\; \Bigg)
 \end{align}
 so for nearly flat models, where $r << 1$,
 \begin{align}
   ({\cal M} - M)/M \approx 0.3 k r^2
 \end{align}
 The pnc and AH equations (\ref{tn_DE_LT}), (\ref{dRndz_LT}), (\ref{dRndz_LT_0}), 
(\ref{AH_cubic_LT}) \& (\ref{dmundz}), using $S_m = S(\hat{t}(r_m))$, become
 \begin{align}
   \hat{t}' = \frac{- S}{\sqrt{1 - k r^2\;}} &
      ~, \\
   \td{R_n}{z} = \left[ \frac{\sqrt{1 - k r^2}\; - r \dot{S}}{H (1 + z)} \right]_n & ~,~~~~
   \td{R_n}{z} \Bigg|_{z = 0} = \frac{1}{H_0} ~,
   \label{dRndz_RW} \\
   \Lambda r_m^3 \hat{S}_m^3 - 3 r_m \hat{S}_m  & + 6 M_0 r_m^3 = 0 ~,
   \label{AH_cubic_RW} \\
   \frac{6 M_0}{\kappa \dot{S}^3 (1 + z)^2} \Big( 2 r \sqrt{1 - k r^2}\; \dot{S} & 
      + r^2 (S \ddot{S} - \dot{S}^2) \Big) = 0
   \label{mun_min_RW}
 \end{align}
 Figs \ref{R-zCurves} and \ref{mun-zCurves}, show $\hat{R}(z)$ and $\mu n(z)$ for a 
range of $H_0$, $\Omega_\Lambda$ and $\Omega_m$ values.  The primary interest is on 
how these parameters affect the maxima.

 \end{document}